\documentstyle[12pt,epsfig,amssymb]{article} 

\begin{document}
\title{
\vspace*{-1.5cm}
\begin{flushright}
{\normalsize DO--TH 98/09\\[1.0cm]}
\end{flushright}
New Analysis of the $\Delta I = 1/2$ Rule
in the \\
$1/N_c$ Expansion 
for $K \rightarrow \pi \pi$ Decays
\thanks{Presented 
at the XXIth School of Theoretical Physics "Recent Progress 
in Theory and Phenomenology of Fundamental Interactions",
Ustro\'n, Poland, September 19-24, 1997, published in 
Acta Phys. Pol. {\bf B28}, 2479 (1997). Two typos in
Eq.~(17) and Eq.~(18) have been corrected.}
}
\author{Thomas Hambye\\
\vspace{3mm}
\normalsize{\it Dept. of Physics (T3),
Dortmund University, 44221 Dortmund, Germany}}
%}
\date{}
\maketitle
\begin{abstract}
We analyze long-distance contributions to the $K \rightarrow \pi \pi$
amplitudes relevant for the
$\Delta I = 1/2$ selection rule 
in the framework of the $1/N_c$
expansion. We use a modified prescription for the identification of
meson momenta in the chiral loop corrections to gain a consistent
matching with the short-distance part. Our approach involves a separation
of non-factorizable and factorizable $1/N_c$ corrections.
Along these lines we calculate the one-loop contributions from the
lowest order lagrangian. Our main result is an additional enhancement
of the $\Delta I = 1/2$ channel amplitude 
which we find in good
agreement with experiment.
\end{abstract}
%\PACS{12.38.Lg;12.39.Fe;13.20.Es;14.40.Aq}
\section{Introduction}
Since the first observation of the $\Delta I = 1/2$ enhancement more than 40
years ago \cite{GP}, there have been many attempts to find dynamical
mechanisms responsible for it, in particular within the standard model.
This $\Delta I =1/2$ rule was particularly 
enigmatic before the birth of QCD 
when 
only the current-current operator $Q_2$ arising from the W exchange
was considered
and, consequently, the ratio $ReA(K \rightarrow (\pi \pi)_{
I = 0})/$$ReA(K \rightarrow (\pi \pi)_{I=2})$$\equiv
ReA_0/ReA_2\equiv R$ was expected to be
about one order of magnitude smaller than
the experimentally observed value $R = 22.2$. Now, since the
esta\-blish\-ment of QCD, our understanding of this rule has improved
considera\-bly. Within QCD, the $K \rightarrow \pi \pi$ amplitudes are
obtained from the effective hamiltonian for
$\Delta S =\nolinebreak 1$ transitions \cite{GLAM,VSZ,GWGP},
\begin{equation}
{\cal H}_{ef\hspace{-0.5mm}f}^{\scriptscriptstyle \Delta S=1}=
\frac{G_F}{\sqrt{2}}
\;V_{us}^* V_{ud}^{}\sum_{i=1}^8 c_i(\mu)Q_i(\mu)\;,
\label{Heff}
\end{equation}
involving the Wilson coefficients $c_i(\mu)$
which can be calculated for a scale
$\mu \gtrsim 1$ GeV using 
perturbative renormalization group techniques, as well as,
the local four-quark 
operators $Q_i(\mu)$. The hadronic matrix elements
of these operators are difficult
to calculate but can be estimated using non-perturbative techniques
generally for $\mu$ around a scale of $1$ GeV. For the $\Delta I = 1/2$
rule only the $z_i$ part 
of the Wilson coefficient $c_i$ is numerically relevant,
with $c_i(\mu)=z_i(\mu)-y_i(\mu)V_{ts}^*V_{td}^{}/(V_{us}^*
V_{ud}^{})$. 
The dominant operators are 
\begin{equation}
Q_1=4 \bar{s}_{\scriptsize L} \gamma^\mu d_{\scriptsize L} 
\bar{u}_{\scriptsize L}
 \gamma_\mu u_{\scriptsize L}\;,
\hspace{2mm}
Q_2=4 \bar{s}_{\scriptsize L} \gamma^\mu u_{\scriptsize L} 
\bar{u}_{\scriptsize L} \gamma_\mu 
d_{\scriptsize L}\;,
\hspace{2mm}
Q_6 =-8\hspace{-2mm}\sum_{q=u,d,s}\hspace{-2mm}
\bar{s}_{\scriptsize L} q_{\scriptsize R}\,\bar{q}_{\scriptsize R} 
d_{\scriptsize L} \;,
\label{Qis}
\end{equation}
with $q_{\scriptsize R,L}=\frac{1}{2}(1 \pm \gamma_5)q$.
Major improvements were obtained when it was observed that the QCD
(and electroweakly) induced 
effective hamiltonian of Eq.(\ref{Heff}) can explain various
large enhancements. These can be of short-distance (SD) nature,
like the first identified {\it octet enhancement} \cite{GLAM} 
in the $Q_1$-$Q_2$
sector dominated by the increase of $z_2$ when $\mu$ evolves
from $M_W$ 
down to $\mu \simeq 1$ GeV. Another important SD enhancement
was found
to arise in the sector of the 
QCD penguin operators, in particular for $z_6$, 
through the proper inclusion of the treshold effects (and the
associated incomplete GIM cancellation)
\cite{BBG}. Enhancements are also of long-distance (LD)
nature like the 
first identified LD enhancement of the matrix elements of 
the QCD penguin operators over the matrix elements of $Q_1$ and $Q_2$.
The latter was 
first conjectured and estimated in Ref.~\cite{VSZ} 
using the vacuum insertion method. 
Due to the non-perturbative character 
of the LD contribution,
a large variety of techniques has been proposed to estimate
it
(for some recent publications see Ref.~\cite{Var}). 
Among 
the analytical methods, the $1/N_c$ approach
\cite{BBG} associated with the chiral effective
lagrangian 
is particularly interesting. 
In this approach, an additional LD enhancement
is obtained from inclusion of
chiral loop effects in the $Q_1$-$Q_2$ sector.
The net result of all enhancements mentioned above is a value of $R$
in the range of $70$-$75 \%$ \cite{BBG} of the 
measured value $R=22.2$,
suggesting that the bulk of the $\Delta I = 1/2$ rule in $K
\rightarrow \pi \pi$ decays is now understood.\footnote{In fact
the $\Delta I
= 3/2$ channel is usually obtained
"sufficiently suppressed" whereas the $\Delta I = 1/2$ enhancement
is obtained only partially.
In Ref.~\cite{BBG} for $m_s(1$ GeV$) = 150$ MeV and 
$\Lambda_{\mbox{\scriptsize {QCD}}}\equiv \Lambda^{(4)}=300$ MeV 
the latter channel was 
reproduced to $\simeq70 \%$.}
One might note that
the agreement with experiment is not improved by 
inclusion of the NLO renormalization group equations for the $z_i$'s
\cite{BBG2}.

In this proceeding, we reconsider the calculation of 
$K \rightarrow \pi \pi$ amplitudes relevant 
for the $\Delta I = 1/2$ rule in the $1/N_c$
approach of Ref.~\cite{BBG}. Our main  improvement consists in the fact
that we use a modified matching procedure in order to remove
previous ambiguities. This will be done treating the
factorizable (F) and non-factorizable (NF) contributions differently. 
Only the NF diagrams are matched with the SD contributions to cancel
the scale dependence of the SD Wilson coefficients (after having, as
we will see, facto\-rized out
the scale dependence of the coefficient $1/m_s^2$ in the matrix elements of 
the operator $Q_6$).
Factorizable loop contributions which refer uniquely to the strong sector
of the theory can be calculated in full chiral perturbation
theory ($\chi$PT), the corresponding scale dependence being absorbed
in the renormalization of the various parameters in the lagrangian.
As a result of this procedure, performing a matching with the
SD Wilson coefficients, we obtain an additional enhancement of the $\Delta
I = 1/2$ channel which we
find to be in good agreement with experiment. The $\Delta I =
3/2$ channel, however, exhibits a large dependence on the matching scale,
resulting from the  difference of two large numbers, but is found to be
equal or smaller than the experimental value in such a way that the
ratio $R=22.2$ is reproduced or even
passed beyond.
%
%%%%%%%%%%%%%%%%%%%%%%%%%%%%%%%%%%%%%%%%%%%%%%%%%%%%%%%%%%%%%%%%%%%%%%%%
\section{General framework \label{GF}}
To calculate the matrix elements, we start from the chiral 
effective lagrangian for 
pseudoscalar mesons which involves an expansion in momenta where terms up 
to ${\cal O}(p^4)$ are included
\cite{GaL},
\begin{eqnarray}
{\cal L}_{ef\hspace{-0.5mm}f}&=&\frac{f^2}{4}\Big(
\langle \partial_\mu U^\dagger \partial^{\mu}U\rangle
+\frac{\alpha}{4N_c}\langle \ln U^\dagger -\ln U\rangle^2
+r\langle {\cal M} U^\dagger+U{\cal M}^\dagger\rangle\Big) 
\nonumber
\\ && +rL_5\langle \partial_\mu U^\dagger\partial^\mu U({\cal M}^\dagger U
+U^\dagger{\cal M})\rangle
\;,\label{Leff}
\end{eqnarray}
with $\langle A\rangle$ denoting the trace of $A$ and ${\cal M}= 
{\mbox {diag}}(m_u,\,m_d,\,m_s)$. $f$ and $r$ are free parameters 
related to the pion decay constant $F_\pi$ ($=93$ MeV) 
and to the quark condensate, 
respectively, with $r=-2\langle \bar{q}q\rangle/f^2$.
Up to terms of ${\cal O}(p^4)$ and to leading order in the $1/N_c$
expansion, the lagrangian of Eq.(\ref{Leff}) has the most general structure
relevant for the operators $Q_1$, $Q_2$ and $Q_6$.
The degrees of freedom of the complex matrix $U$ are identified with the 
pseudoscalar meson nonet given in a non-linear representation:
\begin{equation}
U=\exp\frac{i}{f}\Pi\,,\hspace{1cm} \Pi=\pi^a\lambda_a\,,\hspace{1cm} 
\langle\lambda_a\lambda_b\rangle=2\delta_{ab}\,, 
\end{equation}
where, in terms of the physical states,
\begin{equation}
\Pi=\hspace{-0.4mm}\left(
\begin{array}{ccc}
\textstyle\pi^0+\frac{1}{\sqrt{3}}a\eta+\sqrt{\frac{2}{3}}b\eta'
& \sqrt2\pi^+ & \sqrt2 K^+  \\[2mm]
\sqrt2 \pi^- & \textstyle
-\pi^0+\frac{1}{\sqrt{3}}a\eta+\sqrt{\frac{2}{3}}b\eta' & \sqrt2 K^0 \\[2mm]
\sqrt2 K^- & \sqrt2 \bar{K}^0 & 
\textstyle -\frac{2}{\sqrt{3}}b\eta+\sqrt{\frac{2}{3}}a\eta'
\end{array} \right)\,,\hspace{3mm}
\end{equation}
with
$a= \cos \theta-\sqrt{2}\sin\theta$ and
$\sqrt{2}b=\sin\theta+\sqrt{2}\cos\theta$.
$\theta$ is the $\eta-\eta'$ mixing angle
satisfying the relation 
\cite{eta}
\begin{equation}
\tan 2\theta=\frac{2m_{80}^2}{m_{00}^2-m_{88}^2}=2\sqrt{2}\left[1-
\frac{3\alpha}{2(m_K^2-m_\pi^2)}\right]^{-1}\,,
\end{equation}
which yields $\theta \simeq -19^\circ$. 
Note that we treat the singlet as 
a dynamical degree of freedom. Consequently, in Eq.(\ref{Leff}) we include 
the strong anomaly term, with the instanton 
parameter $\alpha$ ($\simeq 0.72$ ${\mbox {GeV}}^2$), 
which gives a non-vanishing mass of the $\eta_0$ in the chiral 
limit ($m_q=0$) reflecting the explicit breaking of the axial $U(1)$ symmetry. 
The lagrangian of Eq.(\ref{Leff}) is equivalent to the one of Ref.~\cite{BBG}
(provided we identify the coefficient $1/\Lambda_\chi^2$ of Ref.~\cite{BBG} 
with $4 L_5/ f^2$), except for the fact that we explicitly 
include the $\eta_0$.

A straightforward bosonization yields the chiral representation of the
quark currents and densities
\begin{eqnarray}
(J_L^\mu)_{ij}&=&\bar{q}_{iL} \gamma^\mu q_{jL} =
-i\frac{f^2}{2}(U^\dagger \partial^\mu U)_{ji}+i r L_5
\Big( \partial^\mu U^\dagger {\cal M} 
-{\cal M}^\dagger \partial^\mu U  \nonumber \\
&& \hspace{2.0cm}+\partial^\mu U^\dagger U {\cal M}^\dagger U 
- U^\dagger {\cal M} U^\dagger 
\partial^\mu U\Big)_{ji}\;, \label{CJ} \\[4mm]
(D_L)_{ij}&=&(D_R^\dagger)_{ij}=
\bar{q}_{iR} q_{jL} 
=-r\Big(\frac{f^2}{4}U^\dagger+L_5\partial_\mu U^\dagger
\partial^\mu U U^\dagger 
\Big)_{ji}\;.
\label{CD}
\end{eqnarray}
Using Eqs.~(\ref{CJ})-(\ref{CD}), the operators $Q_1$, 
$Q_2$ and $Q_6$ can be expressed
in terms of the meson fields.

The $1/N_c$ corrections to the matrix elements $\langle Q_i\rangle_I$ are
calculated by chiral loop diagrams. 
The loop expansion involves a series in $1/f^2\sim1/N_c$ which is 
in direct correspondence with the short-distance expansion in terms of 
$\alpha_s/\pi\sim 1/N_c$. 
In these diagrams we encounter 
integrals which are regularized by a finite cutoff as it
was introduced in Ref.~\cite{BBG}. Due to the truncation to pseudoscalar 
mesons, the cutoff has to be taken at or, preferably, 
below the ${\cal O}$(1GeV). 
This restriction is a common feature of the phenomenological approaches at 
hand in which higher resonances are not included. 

To retain the physical amplitudes $A_I$, which as a matter of principle are 
scale-inde\-pendent, the long- {\it and} the short-distance contributions are 
eva\-luated at the cutoff scale, i.e., the LD ultraviolet scale is 
identified with the SD infrared one. Performing this identification
we must take into account that, within the cutoff regularization, there is a
dependence on the way we define the momentum variable inside the loop.

In the standard approach of Refs.~\cite{BBG,JMS1,EAP2}, 
the cutoff is associated 
to the virtual meson, i.e., the integration variable is identified with the 
meson momentum. Consequently, as there is no 
corresponding quantity in the short-distance part, a rigorous matching of 
long- and short-distance contributions is not possible.

The ambiguity is removed by associating the cutoff to the effective color
singlet gauge boson as introduced within a study of the $K_L$--$K_S$ mass 
difference \cite{BGK}.
This is done by identifying in the LD, as well as in the SD part, 
the momentum variable in the loop integration with
the momentum flow between the two currents or densities.
With respect to the standard approach,
the momentum of the virtual meson is shifted by the 
external momentum, the former being no longer identical with the integration 
variable, which affects both the ultraviolet, as well as the infrared 
structure of the $1/N_c$ corrections. Note that the matching
prescription advocated in Ref.~\cite{BGK} was also used for
current-current operators in the chiral limit in Ref.~\cite{FG}.
The authors showed that the coefficient of the quadratic
term in the cutoff 
is increased by a factor 3/2 relative to the
one obtained from the standard matching prescription \cite{BBG,JMS1,EAP2}. 
This provides
an additional {\it octet enhancement}
which,
however, has to be confirmed by a full calculation of the amplitudes
relevant for the $\Delta I = 1/2$ rule. 
As we will argue, from this calculation
we performed, we confirm
this enhancement which is even largely
increased by the effects beyond the chiral limit.

Obviously, the modified procedure described above is applicable  to
the non-factori\-zable part of the interaction
and not to the factorizable ones.
The factorizable part, however, refers
to the strong sector of the theory,
and has not to be matched with any SD contribution.
Consequently, it can be calculated in
pure $\chi$PT. In this case dimensional
regularization can (and will) 
be used. Therefore, no momentum prescription ambiguities appear. This 
separation of F and NF contributions was already
applied in a similar way to investigate the $B_K$ parameter \cite{Bkpar}.

\section{Calculation of the amplitudes and results}

Expanding  the
lagrangian of Eq.(\ref{Leff}), as well as,
the currents and densities of Eqs.(\ref{CJ})-(\ref{CD})
in terms of pseudoscalars fields, 
the various F and NF one-loop diagrams to be calculated are given in Fig.~1
and Fig.~2 respectively. 
\vspace{0.4cm}\\
\noindent
\centerline{\epsfig{file=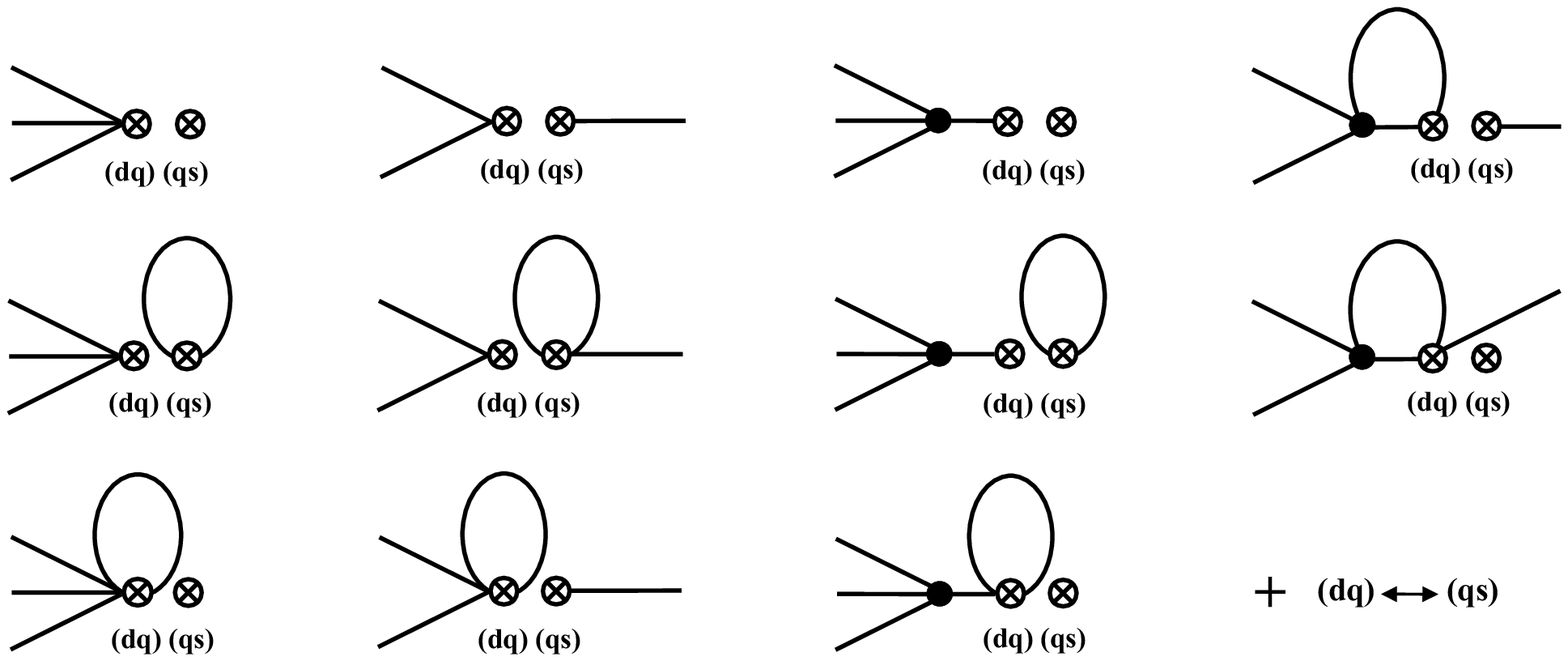,width=10.81cm}}
\\
\vspace{0cm}\\
{\footnotesize Fig.~1. Factorizable diagrams contributing to the matrix 
elements of the operators $Q_i$. Crossed circles 
represent the currents or densities 
(with indices (dq) (qs) here specified for $Q_6$);
black circles denote strong vertices. The lines represent the
pseudoscalar mesons.
}
\vspace{0.3cm}\\

In
addition to these diagrams the effect of the wave function
renorma\-lization must be included, and the values of the
various parameters in the lagrangian must be expressed in terms of
physical quantities. This we did in pure $\chi$PT following
our general procedure for the factorizable contributions.
We
checked that the scale dependence
coming from the factorizable diagrams 
is completely canceled by the scale dependence of the various
parameters in the tree level expression of the matrix elements. 
Explicitly, we obtain the following expressions:
\begin{eqnarray}
i\langle \pi^+ \pi^-|Q_2|K^0 \rangle^F&=&
-i\langle \pi^0 \pi^0 |Q_1|K^0\rangle^F=
X\Big(1+\frac{4L_5^r}{F_\pi^2} m_\pi^2\Big)\;,\label{Fact1}\\
i\langle \pi^+ \pi^-|Q_1|K^0 \rangle^F&=&
i\langle \pi^0 \pi^0 |Q_2|K^0\rangle^F=0\;,\label{Fact2}\\
i\langle \pi^+ \pi^-|Q_6|K^0 \rangle^F&=&
i\langle \pi^0 \pi^0 |Q_6|K^0\rangle^F=
-4 \frac{X}{F_\pi^2}
\bigg(\frac{2m_K^2}{{\hat m}+m_s}\bigg)^2 L_5^r\;,
\label{Fact3}
\end{eqnarray}
with $X=\sqrt{2} F_\pi (m_K^2-m_\pi^2)$ and $L_5^r$ defined by the relation
\begin{equation}
\frac{F_K}{F_\pi}= 1 + \frac{4 L_5^r}{F_\pi^2}(m_K^2 - m_\pi^2)\;.
\label{L5r}
\end{equation}
For the reason of brevity, in Eqs.(\ref{Fact1})-(\ref{Fact3}) we omit
scale-independent (finite) terms resulting from the one-loop corrections
which, nevertheless, are taken into account within the numerical
analysis.
\vspace{1.0cm}\\
\noindent
\centerline{\epsfig{file=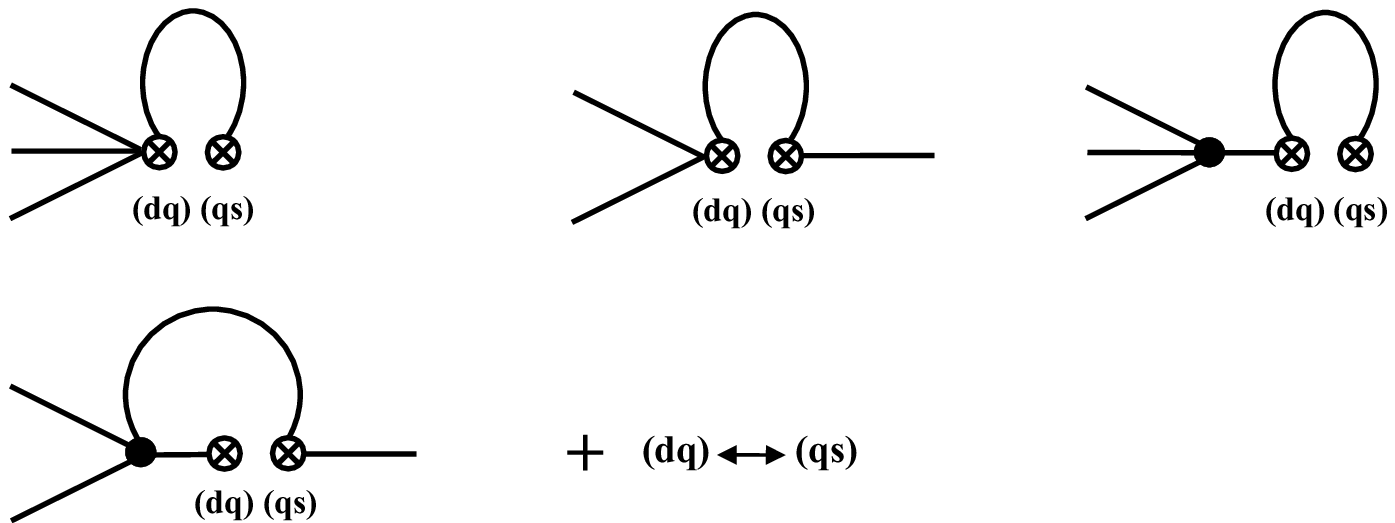,width=8.11cm}}
\\
\vspace{-0.70cm}\\
\begin{center}{\footnotesize Fig.~2. Non-factorizable
diagrams with indices here specified for $Q_6$.
}
\end{center}
\vspace{0.05cm}
Note that in the definition of $L_5^r$, in the denominator on the
left-hand side of Eq.(\ref{L5r}), we used $F_\pi$ rather than f. Formally, 
the difference represents higher order effects. 
Nevertheless,
the appearance of
$f$ in Eq.(\ref{L5r}) would induce residual scale dependence in
Eqs.(\ref{Fact1})-(\ref{Fact3}) which has no counterpart at the SD level.
Therefore, the choice of $F_\pi$
is more adequate as it ensures that no scale dependence
appears in the complete factorizable LD contribution of the matrix
elements except the scale dependence of $m_s$.
To be explicit, in accordance 
with current conservation, there is no scale dependence in
Eqs.(\ref{Fact1})-(\ref{Fact2}); 
as to the density-density operator of Eq.(\ref{Fact3}),
we are just left with the scale dependence of 
$1/m_s^2$.
This is to be expected
since the evolution of $m_s$ is just 
inverse to the evolution of a single density operator.
Consequently, the scale
dependence of Eq.(\ref{Fact3}) (which is of leading order in $N_c$) 
exactly  cancels the 
corresponding diagonal evolution of $z_6$
at SD. This characteristic has already been observed in Ref.~\cite{BBG} .

The non-factorizable contributions are calculated 
from the diagrams of
Fig.~2, 
by means of introducing a cutoff regulator $\Lambda_{NF}$.
Using 
the improved matching prescription as explained above, we obtain
\begin{eqnarray}
i\langle \hspace{0.5mm}\pi^0\hspace{0.5mm}\pi^0\hspace{0.5mm}|Q_1|K^0
\rangle^{NF}&=& 0
\,,\label{NFact1}\\[1mm]
i\langle \pi^+\pi^-|Q_1|K^0
\rangle^{NF}&=& 
\frac{X}{16\pi^2 F_\pi^2}\Big(-3
\hspace{0.6mm}\Lambda_{NF}^2+\frac{1}{4}
\big(m_K^2+12m_\pi^2\big)\ln\Lambda_{NF}^2\Big),
\hspace{5mm}\,\label{NFact2}\\[1mm]
i\langle \hspace{0.5mm}\pi^0\hspace{0.5mm}\pi^0\hspace{0.5mm}|Q_2|K^0
\rangle^{NF}&=& 
\frac{X}{16\pi^2 F_\pi^2}\Big(\hspace{0.5mm}\frac{9}{2}\Lambda_{NF}^2+
\frac{3}{4}\big(m_K^2-6m_\pi^2\big)\ln\Lambda_{NF}^2\Big)
,\hspace{1mm}\,\label{NFact3}\\[1mm]
i\langle \pi^+\pi^-|Q_2|K^0
\rangle^{NF}&=& 
\frac{X}{16\pi^2 F_\pi^2}\Big(\hspace{0.5mm}\frac{3}{2}\Lambda_{NF}^2+
\big(m_K^2-\frac{3}{2}m_\pi^2\big)\ln\Lambda_{NF}^2\Big)
,\hspace{3mm}\,\label{NFact4}\\[1mm]
i\langle \hspace{0.5mm}\pi^0\hspace{0.5mm}\pi^0\hspace{0.5mm}|Q_6|K^0
\rangle^{NF}&=& \frac{X}{16 \pi^2 F_\pi^2}
\frac{3}{4}r^2\ln\Lambda_{NF}^2
,\,\label{NFact5}\\[1mm]
i\langle \pi^+\pi^-|Q_6|K^0\rangle^{NF}&=&
\frac{X}{16 \pi^2 F_\pi^2}\frac{3}{4}r^2 \ln\Lambda_{NF}^2
\,. \label{NFact6}
\end{eqnarray}
Here again we omit
finite terms which, however, are 
taken into account in the
numerical analysis. The scale-dependent 
terms in Eqs.(\ref{NFact1})-(\ref{NFact6}) 
have to be matched
with the Wilson coefficients.
To this end,
we use the numerical values presented in the leading logarithm analyze
of Ref.~\cite{Buch}.

Note that in the denominators of Eqs.(\ref{NFact1})-(\ref{NFact6}) we took
the physical value $F_\pi$ rather than $f$, in the same way 
as for the factorizable diagrams.
Again, the difference
represents higher order effects. However, the scale dependence of $f$ 
in Eqs.(\ref{NFact1})-(\ref{NFact6})
has no counterpart in the SD and will
be absorbed at the next order of the chiral expansion. 
Note also that in Ref.~\cite{BBG} only the tree level contribution 
was taken into account in  the matrix elements of $Q_6$. 
The latter are proportional to the $L_5$ coefficient, as the
tree level contribution of the ${\cal O}(p^0)$ coming from ${\cal L}(p^2)$
vanishes due to the unitarity of $U$.
From the point of the $N_c$ counting it is justified
to consider only the tree level contribution for $Q_6$, 
since the
$z_6$ coefficient is ${\cal O}(1/N_c)$.
In Ref.~\cite{BBG}, only the
one-loop diagrams induced by the operators
$Q_1$ and $Q_2$ were considered. 
Loops over $Q_6$
from ${\cal L}(p^2)$, on the other hand, were assumed to be zero as 
the corresponding tree
level contribution from ${\cal L}(p^2)$ is zero. However, 
an explicit calculation of
the NF one-loop contributions  to the matrix elements of $Q_6$ 
coming from ${\cal L}(p^2)$ yields
a non-trivial cutoff dependence, as shown in Eqs.(\ref{NFact5}) and
(\ref{NFact6}), which 
has to be matched with the SD part. 
This effect cannot be
neglected since it corresponds to the leading non-vanishing order of the
twofold series expansion
in $1/N_c$ and $p^2$;
i.e., it is of the same order as
the tree level term 
proportional to $L_5$ (the former, 
being of ${\cal O}(p^0/N_c)$, is 
the leading term in the $p^2$ expansion, while the latter,
being of  
order ${\cal O}(p^2)$,
is leading in the $1/N_c$ expansion).
The resulting scale dependence, however,
is only of minor importance within the analysis of 
the $\Delta I = 1/2$ rule.\footnote{
It is actually important within the analysis of the ratio
$\varepsilon'/\varepsilon$. Further details on this new calculation of the
matrix elements of $Q_6$ can be found in Ref.~\cite{wip}.}

Our numerical result for the amplitude Re$A_{0}$
is shown in Fig.~3. It shows an additional enhancement
which renders the amplitude in good agreement with
the observed value. The new contribution arises from 
the $Q_1$ and $Q_2$ operators.
It is due to 
the modified matching prescription in the NF sector 
(except for approximately one fifth 
of it which is due to
the choice of the physical value $F_\pi$ in Eqs.(\ref{NFact1})-(\ref{NFact6}) 
as explained above).
Our result is remarkably stable with respect to the matching scale. The
main uncertainty displayed in Fig.~3 originates
from the dependence of the Wilson coefficients 
on $\Lambda_{\mbox{\footnotesize {QCD}}}$ $(\equiv \Lambda^{(4)})$.
On the other hand, the isospin $3/2$  amplitude shown in Fig.~4
is highly unstable. 
The large uncertainty can be understood from the fact that it is
obtained from the difference of two large amplitudes of
approximately the same size [namely
$A(K \rightarrow \pi^+ \pi^-)$ and $A(K \rightarrow \pi^0 \pi^0)$].
Consequently,
the $3/2$ amplitude is not well reproduced
except that, and
this is a crucial point, it comes out to be sufficiently suppressed 
whatever the particular chosen scale is between $500$ MeV
and $1$ GeV.
%\vspace{-2.0cm}\\
\noindent
%\begin{figure}[h]
\protect{\vspace*{15pt}}
\centerline{
%\hspace{+0.26cm}\epsfysize=4.0in {\epsffile{fi2.ps}}
\epsfig{file=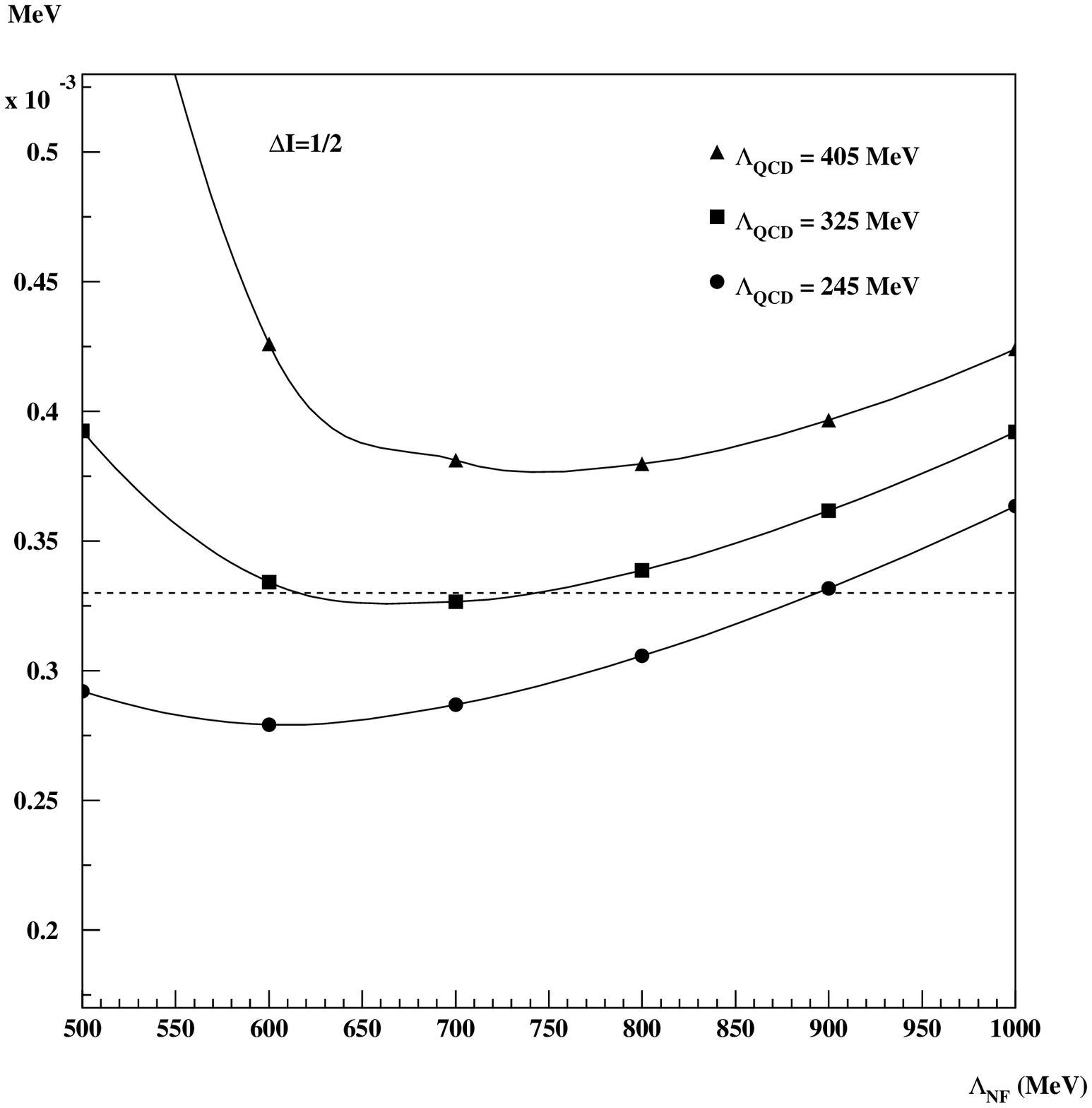,width=8.5cm}}
%}
%\end{figure}
\\
\vspace{-0.4cm}\\
{\footnotesize Fig.~3.  Re$A_0$ in units of $10^{-3}$ MeV 
for $m_s(1$ GeV$)=150$ MeV as a function of the 
matching scale $\Lambda_{NF}$. The experimental 
value is represented by the dashed line.
}
\vspace{-0.1cm}\\

\noindent
%\begin{figure}[h]
\vspace*{10pt}
\centerline{
%\hspace{+0.26cm}\epsfysize=4.0in {\epsffile{fi3.ps}}
\epsfig{file=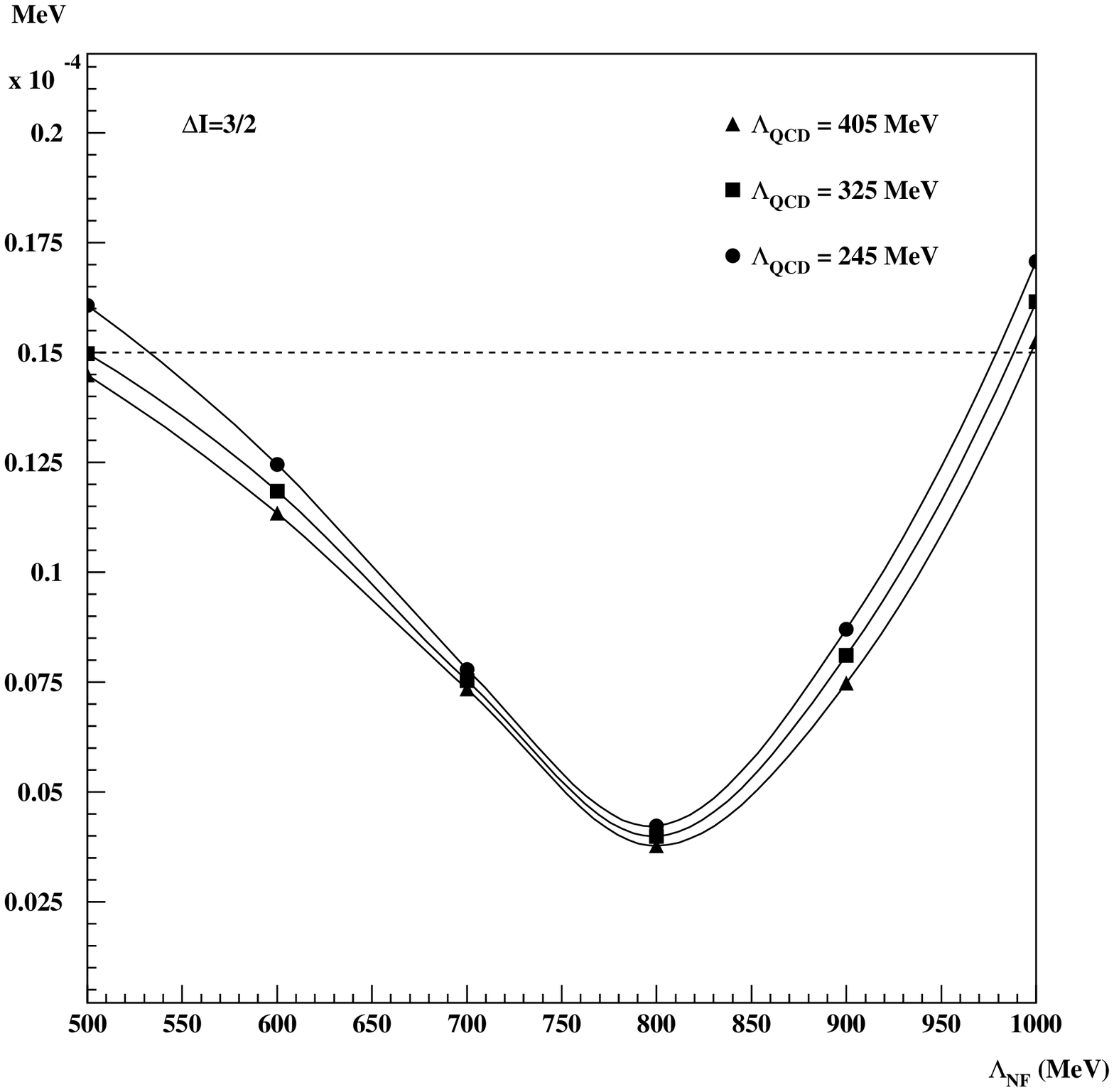,width=8.5cm}}
%}
%\end{figure}
\\
\vspace{-0.9cm}\\
%\centerline{\epsfig{file=fi3.ps,width=6.01cm}}
%
{\footnotesize Fig.~4. 
Re$A_2$ in units of $10^{-3}$ MeV 
for $m_s(1$ GeV$)=150$ MeV.
The experimental 
value is represented by the dashed line.
}
\vspace{0.1cm}\\

In conclusion, it is certainly premature to 
say that the $\Delta I = 1/2$ rule
for $K \rightarrow \pi \pi$ decays is now understood completely since
the $1/N_c$
approach we use is only an approximate method. In particular,
vector mesons and higher resonances should be included in order 
to take a higher and more secure value for the matching scale. 
(It is probable that the vector mesons play an important role
for $A_2$). 
Nevertheless, we believe that
the enhancement
reported here is a further important indication that
the $1/N_c$ approach can account for the bulk
of the $\Delta I = 1/2$ rule for
$K \rightarrow \pi \pi$ decays.

\vspace{0.5cm}
This work has been done in collaboration with G. K\"ohler, E.A. Paschos and
P. Soldan.
We wish to thank W.A. Bardeen, J. Bijnens, J.P. Fatelo  and J.-M. G\'erard for 
helpful comments. The author thanks the Deutsche 
For\-schungs\-gemeinschaft for a postdoctoral fellowship in the 
Graduate Program for 
Elementary Particle Physics at the University of Dortmund.
%

%

%%%%%%%%%%%%%%%%%%%%%%%%%%%%%%%%%%%%%%%%%%%%%%%%%%%%%%%%%%
%%%%%%%%%%%%%%%%%%%%%%%%%%%%%%%%%%%%%%%%%%%%%%%%%%%%%%%%%%%%%%%%%%%%%%%%

\begin{thebibliography}{99}
%
\bibitem{GP}
M. Gell-Mann and A. Pais, Proceeding 
Glasgow Conf. 1954, Pergamon London, 342 (1954).
\bibitem{GLAM}
M.K. Gaillard and B.W. Lee, Phys. Rev. Lett. {\bf 33}, 108 (1974);
G. Altarelli and L. Maiani, Phys. Lett. {\bf B52}, 351 (1974);
\bibitem{VSZ}
M.A Shifman, A.I Vainshtein and V.I Zakharov, JETP {\bf 45}, 670 (1977).
\bibitem{GWGP}
F.J. Gilman and M.B. Wise, Phys. Rev. {\bf D 20}, 2392 (1979);
B. Guberina and R.D. Peccei, Nucl. Phys. {\bf B 163}, 289 (1980). 
%
\bibitem{BBG}
W.A. Bardeen, A.J. Buras, and J.-M. G{\'e}rard, Nucl. Phys. {\bf B293},
787 (1987); Phys. Lett. {\bf B192}, 138 (1987);
A.J. Buras, in {\it CP Violation}, ed. C. Jarlskog, 
World Scientific, 575 (1989).
%
\bibitem{Var}
A. Pich and E. de Rafael, Nucl. Phys. {\bf B358}, 311 (1991);
M. Neubert and B. Stech, Phys. Rev. {\bf D44}, 775 (1991);
M. Jamin and A. Pich, Nucl. Phys. {\bf B425}, 15 (1994);
J. Kambor, J. Missimer and D. Wyler, Nucl. Phys. {\bf B346}, 17 (1990);
S. Bertolini et al.,hep-ph/9705244;
C. Dawson et al., hep-lat/9707009.
%, J.O. Eeg, M Fabbrichesi and E.I. Lashin, hep-ph/9705244.
\bibitem{BBG2}
G. Buchalla, A.J. Buras and M.E. Lautenbacher, Rev. Mod. Phys. {\bf 68},
1125 (1996).
%
\bibitem{GaL}
J. Gasser and H. Leutwyler, Nucl. Phys. {\bf B 250}, 465 (1985). 
%
\bibitem{eta}
J.-M. G{\'e}rard, Mod. Phys. Lett. {\bf A 5}, 391 (1990).
\bibitem{JMS1}
J.-M. Schwarz, {\it Diploma thesis}, Dortmund 1991; J. Heinrich and J.-M.
Schwarz, {\it Internal Report Univ. of Dortmund} 1992-01;
J. Heinrich, E.A. Paschos, J.-M. Schwarz, and Y.L Wu, Phys. Lett. 
{\bf B279}, 140 (1992).
%
\bibitem{EAP2}
E.A. Paschos, Invited Talk presented at the 27th Lepton-Photon Symposium, 
Beijing, China (August 1995), published in {\it Lepton/Photon Symp.~1995}.
%
\bibitem{BGK}
J. Bijnens, J.-M. G{\'e}rard and G. Klein, Phys. Lett. {\bf B257}, 191 (1991).
%
\bibitem{FG}
J.P. Fatelo and J.-M. G{\'e}rard, Phys. Lett. {\bf B347}, 136 (1995).
%
\bibitem{Bkpar}
J. Bijnens and J. Prades, Nucl. Phys. {\bf B 444}, 523 (1995). 
%
\bibitem{Buch}
G. Buchalla, A.J. Buras, and M.K. Harlander, Nucl. Phys. {\bf B337},
313 (1990).
%
\bibitem{wip}
T. Hambye, G.O. K\"ohler, E.A. Paschos, P.H. Soldan and W.A. Bardeen,
preprint DO-TH 97/28, FERMILAB-Pub-98/051-T (hep-ph/9802300), to appear in
Phys. Rev. {\bf D}.
%\bibitem{PS}
%P.H. Soldan, preprint DO-TH 96/15 (hep-ph/9608281), 
%invited talk presented at the {\it Workshop on K Physics}, Orsay,
%France, May 30 -- June 4, 1996; published in the proceedings. 
\end{thebibliography}
\end{document}